\documentclass{article}

\usepackage{PRIMEarxiv}

\usepackage[utf8]{inputenc} % allow utf-8 input
\usepackage[T1]{fontenc}    % use 8-bit T1 fonts
\usepackage{hyperref}       % hyperlinks
\usepackage{url}            % simple URL typesetting
\usepackage{booktabs}       % professional-quality tables
\usepackage{amsfonts}       % blackboard math symbols
\usepackage{nicefrac}       % compact symbols for 1/2, etc.
\usepackage{microtype}      % microtypography
\usepackage{lipsum}
\usepackage{fancyhdr}       % header
\usepackage{graphicx}       % graphics
\usepackage{subcaption}
\usepackage{comment}
\graphicspath{{media/}}     % organize your images and other figures under media/ folder

\title{Evaluating Different Modalities of Behavioral Approach Tests for Spider Phobia in Virtual Reality}

\author{
    Florian Grensing \\
    Data Engineering\\
    Helmut-Schmidt-University \\
    Hamburg\\
    \texttt{Grensinf@hsu-hh.de} \\
    \And
    Vanessa Schmücker\\
    Chair of Clinical Psychology\\
    University of Siegen\\
    Siegen, Germany\\
    Vanessa.Schmuecker@uni-siegen.de\\
    \And
    Anne Hildebrand\\
    Chair of Psychiatry and Psychotherapy\\
    Bielefeld University\\
    Bielefeld, Germany\\
    Anne.Hildebrand@evkb.de\\
    \And
    Tim Klucken\\
    Chair of Clinical Psychology\\
    University of Siegen\\
    Siegen, Germany\\
    Tim.Klucken@psychologie.uni-siegen.de\\
    \And
    Maria Maleshkova \\
    Data Engineering\\
    Helmut-Schmidt-University \\
    Hamburg\\
    \texttt{Maleshkm@hsu-hh.de} \\
}

\begin{document}
\maketitle

\begin{abstract}

Behavioral approach tests (BAT) are a common means of assessing specific phobias. In these tests, participants move towards a anxiety-inducing stimulus as close as they are willing to, with the final distance indicating the severity of the anxiety.       

In this work, we aim to evaluate a virtual reality implementation of the BAT. For this purpose, four different BATs were designed, consisting of two approach methods, both replicated in vivo and in virtuo. Evaluation of these BATs are done by using a standardized presence questionnaire, application-specific questions, as well as physiological reactions of the participants. The study focuses on the fear of spiders and uses a real and virtual spider as anxiety-inducing stimulus.  

Our results show that the developed VR BAT perform within established presence norms, while the different modalities influenced participants’ subjective impressions. Furthermore, the standardized structure of the VR environment ensured a consistent experience regarding the anxiety-inducing stimulus. This differs from the observation in the real-world setting, where the behavior of the spider might differ between individuals and also between sessions. This highlights one of the key advantages of virtual reality, which is the complete control of the stimulus and environment. 
Correlations between presence and physiological signals were found. Particularly, tonic electrodermal activity levels are more stable with increased presence. However, more research into this is required, as the effects of anxiety on the physiological signals make the correlations difficult to interpret. 

The evaluation has revealed, which design choices are particularly promising for increasing presence in VR applications, and some which should be avoided. Overall, these results indicates that our VR-based implementation is a promising tool for assessing avoidance behavior for individuals with spider phobia.

\end{abstract}

% keywords can be removed
\keywords{arachnophobia \and BAT \and physiological data \and presence \and phobia \and psychology \and virtual reality}

\section{Introduction}

Specific phobias are among the most common anxiety disorders and are characterized by an intense, often irrational fear of certain situations or objects. One of the most prevalent specific phobias is the fear of animals, such as arachnophobia, the fear of spiders \cite{Becker.2007}. This condition is typically associated with strong avoidance behavior and can lead to significant psychological distress. 

Two well-established methods for assessing such phobias are self-reporting questionnaires and tests like the behavioral approach test (BAT). In this task, individuals are guided by a clinician to approach the feared stimulus as closely as they are willing and able to. The final distance maintained from the fear-inducing stimulus serves as an indicator of the severity of the phobic response. 

In recent years, virtual reality (VR) has gained increasing attention as a complementary or even alternative medium for exposure therapy in the treatment of specific phobias \cite{Miloff.2019,Kuleli.2025,Albakri.2022, Andersson.2024}. VR allows for the controlled, customizable, and location-independent simulation of anxiety-inducing scenarios. Especially in the treatment of arachnophobia, VR enables the creation of various realistic settings, such as basements, attics, or bathrooms. Additionally, virtual reality can replace the need for a living spider, further increasing the flexibility and preventing animal suffering. One important factor influencing the effectiveness of VR-based exposure is the sense of presence, defined as the subjective feeling of ``being there'' in the virtual environment. Studies have shown that higher levels of presence are associated with stronger emotional and physiological responses and may enhance therapeutic outcomes \cite{Augustin.2024,Lemmens.2022}.

This work explores the idea of transferring the benefits of VR-based exposure therapy to the BAT paradigm. 
A VR application consisting of two BAT modalities was developed to assess the fear of spiders, involving either approaching the spider by walking toward it or by actively pulling it closer using a crank mechanism. Additionally, participants completed these tests in two real-world conditions that mirrored the VR tasks, using the same approach modalities. The goal of this paper was to examine potential differences in presence, physiological responses, and subjective evaluations of the scenarios. The findings aim to provide insights into how different forms of virtual confrontation affect individuals with spider phobia and to determine design aspects that are suitable for increasing presence in future applications. 

\section{Research Background}\label{researchBackground}

%Transfer of Exposition into VR
In recent years, a growing field of research has investigated the benefits of virtual reality in the field of psychology, especially regarding virtual reality exposure therapy (VRET). Recent studies, such as those by Miloff et al. \cite{Miloff.2019} and Andersson et al. \cite{Andersson.2024}, suggest that VRET yields outcomes comparable to traditional in vivo exposure. These results suggest that VR-based exposure therapy is a feasible and effective alternative, especially considering its potential for standardization, accessibility, and user control.

%Transfer of BAT to VR
Previous studies in this field hint that the BAT can also be applied to virtual reality. According to Mühlberger et al. \cite{Mühlberger}, the approach and anxiety measures correlated with psychometric measures of spider phobia. However, this and similar studies do not compare the effectiveness of BAT performed in virtuo and in vivo. Likewise, the expectations of the participants and the presence experienced in VR, which are particularly addressed in our work, are not recorded either.

%What are BAT
As previously mentioned, a BAT consists of participants approaching a feared stimulus as closely as possible. However, different methods of approaching this feared stimulus exist. Mühlberger et al. \cite{Mühlberger} describe a setup, in which the participant enters a room and then walks towards the spider. Schwarzmeier et al. \cite{Schwarzmeier.2020} on the other hand used a setup, in which the participants pull the spider towards themselves using a crank-like device. During the test, participants can be tasked with describing their current anxiety on a scale of 0 (no anxiety) to 100 (extreme anxiety). These subjective anxiety ratings, as well as the final distance to the stimulus are then used as the outcome of the BAT.

%Presence important
For transferring the BAT into virtual reality, the effect of presence also becomes relevant. Presence refers to the immersive experience in VR, whereby the user is engaged as strongly as if the virtual world were real. This is important, as an increased presence leads to a better transfer of the learned behavior into real life, thus improving treatment outcomes. This can be seen in works that show a positive correlation between the sense of presence and emotional responses \cite{MatthewPrice.2007, Ling.2014}. Another aspect of presence is shown in the physiological response. Past works, such as the one by Frieden et al. \cite{PräsenzVital} show a correlation between presence and physiological reactions, particularly in electrodermal activity (EDA). 

%Summary and research gap
Despite the growing interest in this field, there is a notable lack of recent publications addressing the direct transfer of the behavioral approach test into virtual reality environments. In many studies, a traditional in vivo BAT is still used solely as an outcome measure to verify the success of VRET interventions. This reveals a gap in the current research landscape, that this study seeks to address by exploring the implementation and evaluation of two fully virtual BAT setting.

\section{Behavioral Approach Test Scenarios}

The behavioral approach test is a widely used method in clinical psychology to assess anxiety and monitor therapeutic progress. In this study, we conducted the BAT both in vivo and in virtual reality, as well as comparing two distinct approach modalities. The two approach modalities are based on the works of Schwarzmeier et al. \cite{Schwarzmeier.2020} and Mühlberger et al. \cite{Mühlberger}. To ensure optimal comparability, both were replicated into virtual reality. To this end, four experimental conditions were defined:

\begin{itemize}
    \item VRT: Tutorial for participants to learn the controls in VR. 
    \item VIVO1: The participant cranks the anxiety-inducing stimulus toward themselves in a real-world setting.
    \item VR1: The participant cranks the anxiety-inducing stimulus toward themselves in VR.
    \item VIVO2: The participant walks toward the anxiety-inducing stimulus in a real-world setting.
    \item VR2: The participant walks toward the anxiety-inducing stimulus in VR.
\end{itemize}

%Created in collaboration
These scenarios were developed through an interdisciplinary collaboration between the former Chair of Medical Informatics and the Chair of Clinical Psychology at the University of Siegen, as well as the Chair of Data Engineering at the University of the Federal Armed Forces in Hamburg.
The target group of the application are spider phobics, whose avoidance behavior is to be measured by the different scenarios. For this purpose, a 3D model of a tarantula was used, which is very similar to the real spider in vivo. 

%Virtuo created to be similar to vivo counterparts
To ensure ideal replicable VR conditions, the HTC VIVE Lighthouse system \cite{VivePro} was used, which allows for room-scale tracking and unrestricted movement in virtual environments. This technology enabled the virtual scenario in VR2 to closely mirror the real-world setup of VIVO2, including the spatial layout and the walkable area. Consequently, participants had comparable freedom of movement in both VR and in vivo conditions when approaching the anxiety-related stimulus.

%Difference in how anxiety in reported
One notable difference, however, was the method of reporting anxiety. In the in vivo conditions (VIVO1 and VIVO2), anxiety intensity was reported verbally by the participants, whereas in the VR conditions, anxiety was recorded digitally via a virtual slider embedded in the application. This design choice enabled the precise capture of anxiety responses directly within the VR system and opens the possibility for future adaptive VR systems that respond dynamically to users’ anxiety levels. Voice recognition was considered but discarded due to concerns regarding reliability and environmental noise.

\section{VR Application}
The implemented VR application, consisting of both modalities, and the study design have already been presented in detail in another of our papers \cite{BATConcept}. Therefore, this paper will only briefly discuss the implementation and design. The VR BAT was developed in the Unity \cite{Unity} programming environment (2021.3.2f1) and was optimized for the HTC VIVE Pro VR system.
The VR application consists of three scenarios. The first is a tutorial, which is used to practice the BAT procedure and the use of the virtual slider to enter anxiety ratings. For this purpose, a slice of cake must be moved towards the user by the touchpad of the VR controller. Before the start, at the start and for every 25\% of total distance traveled, anxiety ratings are given on a scale from 0 (no anxiety) to 100 (extreme anxiety). These anxiety ratings are also given in each of the BATs, using the same distances and scale.

\begin{figure}[t]
    \centerline{\includegraphics[width=0.5\textwidth]{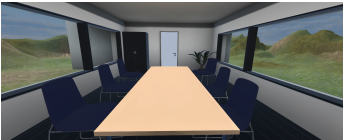}}
    \caption{View of the Crank setup of VR1 \cite{BATConcept}}
    \label{fig:VR1}
\end{figure}

VR1 is similar to the tutorial, but includes some additional steps, based on VIVO1. These additional steps are a virtual study supervisor carrying the spider into the room and placing it on the platform of the crank device. This is required, as an anxiety rating has to be given as the supervisor carries the spider into the room in both VR1 and VIVO1. Once the spider is on the virtual table, it is pulled towards the user using the controller, as the cake in the tutorial. However, while this is a replication of the crank system used in VIVO1, a visual representation of the crank system has been omitted to avoid potential distractions and keep the focus of the user on the spider. 
The table and environment of VR1 can be seen in ``Fig.~\ref{fig:VR1}''.

\begin{figure}[t]
    \centerline{\includegraphics[width=0.8\textwidth]{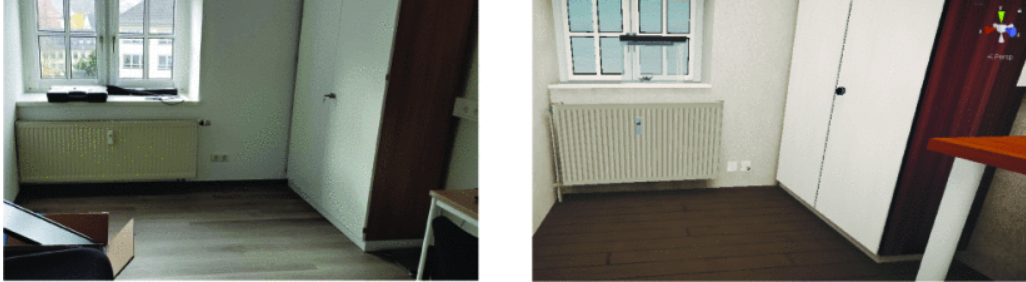}}
    \caption{Side by side comparison of VIVO2 and VR2 \cite{BATConcept}}
    \label{fig:VR2}
\end{figure}

VR2 uses the same anxiety input system. However, as no interaction between the supervisor and the spider or participant is required, no virtual supervisor is implemented. Instead, the spider is placed on the windowsill on the far end of the room before the BAT takes place. The BAT begins when the door to the room is opened. In both VR2 and VIVO2, the participants enter the room and move freely towards the spider. At the aforementioned intervals, they are prompted to provide an anxiety rating, either by a controller vibration in VR2, or orally in VIVO2. The room was recreated in VR to be able to, specifically, analyze possible influences of the realistic environment on the presence. An example of the input method can be seen inA side by side comparison of the real and virtual environments for VIVO2 and VR2 can be seen in ``Fig.~\ref{fig:VR2}''. 

All scenarios can be canceled by the supervisor at any time, at the user's request. There is a corresponding button on the PC interface for this purpose, whereby the procedure ends after a final anxiety rating is given at the final position so that the user can be quickly released from the anxiety-inducing situation. 

\section{Study Design}

The study started in June 2022 and included 25 participants, consisting of 3 (12\%) males and 22 females (88\%), with a mean age of 24.96 years (SD = 8.75). Participants underwent screening for spider phobia before the start of the study, as an active spider phobia was an inclusion criteria. The research was approved by the Ethics Committee of the University of Siegen (reference number: ER\_04\_2023).

At the start of the study, the participants were asked to sign a consent form, after which the Empatica E4 was attached to their wrists to measure physiological data. The study began with the VR tutorial to familiarize participants with the virtual world. After completing this task, the users were asked to fill in a set of questionnaires. Then, one of the BAT scenarios was started in pseudo-randomized order. This was achieved by using a balanced randomization where each of the possible sequences of BATs were performed at least once.
Each BAT ran until the participants were unable or unwilling to continue or reached the spider. After each test, multiple questionnaires related to the performed BAT was filled out. When the questionnaires were filled out, and if the stress levels have returned to baseline, the participants were led to the next BAT. This process was then repeated until the participants have performed all four variants.

%VR Presence questionnaires
In order to evaluate the immersive experience of both VR implementations and the tutorial, the \textit{Presence Questionnaire} originally developed by Witmer and Singer and later revised by the UQO Cyberpsychology Lab \cite{Witmer.2005} was utilized. This revised questionnaire consists of 24 items, which are grouped into eight presence-related dimension: \textit{Overall Presence}, \textit{Realism}, \textit{Possibility to Act}, \textit{Quality of Interface}, \textit{Possibility to Examine}, \textit{Self-Evaluation of Performance}, \textit{Haptic} and \textit{Sound}.
Each item was rated on a 7-point Likert scale (ranging from 1 to 7). To contextualize the findings, the resulting values were compared with norm values provided by the UQO Cyberpsychology Lab. Note that no normative values were available for the haptic feedback and sound dimensions. 

In addition to presence measures, further application-specific questions were posed regarding the impression of the spider, the design of the application, and its usability. Responses were collected using a Likert scale from 0 to 6, where 0 indicated no agreement and 6 indicated full agreement.

%Physiological data & FRAME
Furthermore, physiological data were collected using the Empatica E4 wristband \cite{Empatica}, which measures heart rate, inter-beat interval, skin conductance, and body temperature. These biometric signals allow for quantitative comparison across conditions. The technical foundation for the recording, synchronization and preprocessing of the physiological and observed data are described in our previous works \cite{ReBAT}. Additionally, an analysis of the psychometric data confirmed that BATs performed in VR are indeed a reliable alternative to classical methods \cite{BATAnne}. As such, this work will focus on the evaluation of presence for the VR implementations. 

During the analysis, when determining similarity, we investigate if the data follows a normal distribution. If so, a paired t-test was performed. When normality was not met, a paired Wilcoxon signed-rank was applied instead.

The study session also included additional measurements that were not part of the present research question and are therefore not reported in detail here, as they are not expected to have influenced the recorded parameters \cite{BATAnne}. %The implementation \cite{BATConcept} and participant selection \cite{BATAnne} are described in previous works.

\section{Results}

%For this work, we compare the results between the different modalities for VR1, VR2 and the tutorial for both the presence questionnaires, as we

In this work, the evaluation was split into multiple parts. First, we evaluate the sense of presence experienced by users, reported using the presence questionnaire. Afterwards, we investigate application specific components, such as the behavior of the virtual spider. Lastly, we calculate correlations between physiological reactions and questionnaire outcomes.

\subsection{Presence}

%First, we will evaluate the presence categories, and compare the results to the norms described by the UQO Cyberpsychology Lab. The results are shown in ``Tab.~\ref{tab:vr-comparison}'' and ``Tab.~\ref{tab:vr-categories}'', with no comparison of sound and haptics to the norm, as no norm is described for these categories. 
%Across all categories, the results are within the normative range for all three VR variants. 

First, we will evaluate the presence categories of the Presence Questionnaire, and compare the results to the norms described by the UQO Cyberpsychology Lab. The results show that for all three VR tasks, the presence ratings in all categories are within the specified norms. This can be seen in ``Tab.~\ref{tab:vr-comparison}'' with no comparison of sound and haptic to the norm, as no norm is described for these categories. In addition, mean values and standard deviations can be seen in ``Tab.~\ref{tab:vr-categories}'', which were used to make comparisons between the VR variants within the categories.

\begin{table}[b]
\centering
\scriptsize
\resizebox{\linewidth}{!}{%
\begin{tabular}{l r r r r r r r r}
\toprule
\textbf{Category} & \textbf{Norm Mean} & \textbf{Norm SD} & \textbf{VR1} & \textbf{VR1} & \textbf{VR2} & \textbf{VR2} & \textbf{VRT} & \textbf{VRT} \\
\hline
Total                          & 104.39 & 18.99 & 98.32 & -0.32 & 100.32 & -0.21 & 92.60 & -0.62 \\
Realism                        & 29.45  & 12.04 & 33.36 &  0.32 & 35.72 &  0.52 & 32.52 &  0.25 \\
Act             & 20.76  &  6.01 & 21.16 &  0.07 & 21.12 &  0.06 & 19.88 & -0.15 \\
Interface           & 15.37  &  5.15 & 17.20 &  0.36 & 17.20 &  0.36 & 16.40 &  0.20 \\
Examine         & 15.38  &  4.90 & 15.00 & -0.08 & 15.16 & -0.04 & 13.04 & -0.48 \\
Performance & 11.00  &  2.87 & 11.60 &  0.21 & 11.12 & 0.04 & 10.76 & -0.08 \\ \bottomrule \\
\end{tabular}%
}
\caption{Comparison of dimensions between standard and VR}
\label{tab:vr-comparison}
\end{table}

% \usepackage{booktabs}
\setlength{\tabcolsep}{4pt} % Standard ist meist 6pt
\begin{table}[]
\begin{tabular}{@{}lcccccc@{}}
\toprule
\textbf{Category} & \textbf{VR1 M} & \textbf{VR1 SD} & \textbf{VR2 M} & \textbf{VR2 SD} & \textbf{VRT} & \textbf{VRT SD} \\ \midrule
Total             & 98.32          & 17.72           & 100.32         & 17.80           & 92.60        & 13.60           \\
Realism           & 33.36          & 8.06            & 35.72          & 8.39            & 32.52        & 6.22            \\
Act               & 21.16          & 3.31            & 21.12          & 4.83            & 19.88        & 3.10            \\
Interface         & 17.20          & 2.84            & 17.20          & 3.23            & 16.40        & 3.19            \\
Examine           & 15.00          & 3.16            & 15.16          & 3.95            & 13.04        & 3.16            \\
Performance       & 11.60          & 1.71            & 11.12          & 1.92            & 10.76        & 1.98            \\
Sound             & 14.64          & 4.50            & 13.42          & 5.29            & 14.32        & 3.48            \\
Haptics           & 3.96           & 2.01            & 4.00           & 1.96            & 4.16         & 1.70            \\ \bottomrule \\
\end{tabular}
\caption{Comparison of categories of all VR modalities}
\label{tab:vr-categories}
\setlength{\tabcolsep}{6pt} % zurücksetzen
\end{table}

%Realism => VR2 > VR1 und Tutorial
The \textit{Realism} rating describes how realistic and credible the application appears to the user. A paired-\textit{t}-test revealed that VR2 was rated significantly higher than VR1, $t(24) = 2.09, p = .047, 95\% CI [0.03, 4.69], d = .42, 95\% CI [0.00, 0.82]$, with an average difference of 2.36 points. Compared to the tutorial, this difference was even greater, $t(24) = 2.33, p = .029,95\% CI [0.37,  6.04], d = 0.47, 95\% CI[0.05, 0.87]$ with a mean difference of 3.20 points.

%\paragraph{Possibility to Act}

The \textit{Possibility to Act} indicates the ability to interact with the VR environment. A paired Wilcoxon signed-rank test was applied, but showed no significant difference between VR1 and VR2, $W = 123.5$, $p = .667$, $r = 0.09$, nor between VR2 and the tutorial, $W = 90$, $p = .148$,$r = .297$. In contrast, a paired-samples \textit{t}-test revealed that VR1 scores were significantly higher than tutorial scores, $t(24) = 2.26$, $p = .033$, $95\% CI [0.11, 2.45]$, $d = 0.45$, $95\% CI [0.04, 0.86]$.

%The ability to interact strongly influences the VR experience. VR1 and VR2 yielded similar mean scores of 21.16 (SD = 3.31) and 21.12 (SD = 4.83), respectively. The tutorial received a slightly lower mean score of 19.88 (SD = 3.10). All scores were within the normative range, with VR1 and VR2 closely matching the reference value. As the VR2 data were not normally distributed, a paired Wilcoxon signed-rank test was applied. The test showed no significant difference between VR1 and VR2 ($W = 123.5, p = .667, r = 0.09$), or between VR2 and the tutorial ($W = 90, p = .148, r = .297$). However, there was a significant difference between VR1 and the tutorial, $t(24) = 2.26, p = .033, 95\% CI [0.11, 2.45], d = 0.45, 95\% CI [0.04, 0.86]$, with VR1 scoring on average 1.28 points higher.\\

%\paragraph{Quality of Interface}

The \textit{Quality of the Interface} reflects participants’ perceptions of the VR equipment, including display resolution and input methods. No significant differences were found here between conditions.

%The quality of the interface reflects participants’ perceptions of the VR equipment, including display resolution and input methods. VR1 and VR2 both received a mean score of 17.20 (SD = 2.84 and SD = 3.23, respectively), while the tutorial was rated slightly lower with a mean of 16.40 (SD = 3.19). All scores were within the normative range, and no significant differences were found between conditions. \\

%\paragraph{Possibility to Examine}
In addition to interaction, the \textit{Possibility to Examine} objects within the simulated environment is an important factor contributing to a realistic VR experience. A paired \textit{t}-test confirmed no significant difference between VR1 and VR2 ($t(24) = 1.18$, $p = .748$, $95\% CI[0.856, 1.18]$, $d = 0.07$, $95\% CI[0.33, 0.46]$). Both applications, however, differed significantly from the tutorial: VR1 scored 1.96 points higher ($t(24) = 3.99$, $p < .001$,$95\% CI[0.95, 2.97]$, $d = 0.80$, $95\% CI [0.34, 1.24]$), and VR2 scored 2.12 points higher ($t(24) = 3.63$, $p = .001$, $95\% CI[0.91,3.33]$, $d = 0.73$, $95\% CI [0.28, 1.16]$).

%In addition to interaction, the ability to examine objects within the simulated environment is an important factor contributing to a realistic VR experience. VR1 (M = 15.00, SD = 3.16) and VR2 (M = 15.16, SD = 3.95) again received very similar ratings, whereas the tutorial scored lower with a mean of 13.04 (SD = 3.16). A \textit{t}-test of the normally distributed data confirmed no significant difference between VR1 and VR2 ($t(24) = 1.18, p = .748,95\% CI[0.856, 1.18], d = 0.07, 95\% CI[0.33, 0.46]$). Both applications, however, differed significantly from the tutorial: VR1 scored 1.96 points higher ($t(24) = 3.99, p < .001\% CI[0.95, 2.97], d = 0.80, 95\% CI [0.34, 1.24]$), and VR2 scored 2.12 points higher ($t(24) = 3.63, p = .001\% CI[0.91,3.33], d = 0.73, 95\% CI [0.28, 1.16]$). Despite these differences, all values remained within the normative range.\\

%\paragraph{Self-evaluation of Performance}
The \textit{Self-Evaluation of Performance} assesses how quickly and effectively participants adapted to the new environment and the extent to which they acquired new skills. There was no significant difference between the tutorial and VR2, $t(24) = 1.04$, $p = .30$, $ 95\% CI[-0.35 , 1.07]$, $d = 0.21$, $95\% CI [-0.19 , 0.60]$. However, VR1 differed nearly significantly from VR2, $t(24) = 2.01$, $p = .056$, $95\% CI [-0.01 , 0.97]$ , $d = 0.40$, $95\% CI [-0.01 , 0.81]$, and the difference between VR1 and the tutorial was almost highly significant, $t(24) = 2.58$, $p = .016$, $95\% CI[0.16 , 1.51]$, $d = 0.52$, $95\% CI [0.09 , 0.93]$.

%The self-evaluation of performance assesses how quickly and effectively participants adapted to the new environment and the extent to which they acquired new skills. The ratings were similar across conditions and all within the normative range. Participants reported a mean score of 11.60 (SD = 1.71) for VR1, 11.12 (SD = 1.92) for VR2, and 10.76 (SD = 1.98) for the tutorial. There was no significant difference between the tutorial and VR2 ($t(24) =  1.04, p = .30 95\% CI[-0.35, 1.07], d = 0.21, 95\% CI [-0.19, 0.60]$). However, VR1 differed nearly significantly from VR2 ($t(24) = 2.01, p = .056\% CI[-0.01,0.97], d = 0.40, 95\% CI [-0.01, 0.81]$), and the difference between VR1 and the tutorial was almost highly significant ($t(24) = 2.58, p = .016 95\% CI[0.16, 1.51], d = 0.52, 95\% CI [0.09, 0.93]$).\\

%\paragraph{Sound}
%Sound was not part of the normative framework and thus cannot be compared to the reference values from the UQO Cyberpsychology Lab. A maximum of 21 points could be achieved in this category. The custom applications received very similar ratings: VR1 scored a mean of 14.64 (SD = 4.50), VR2 13.42 (SD = 5.29), and the tutorial 14.32 (SD = 3.48). 
\textit{Sound} describes the auditory impressions of VR.
A maximum of 21 points could be achieved in this category. No significant differences were found between the conditions, indicating only minor variations in sound ratings.

%\paragraph{Haptic}
%Haptic perception was enhanced through vibrations. As with sound, there was no normative reference, and the maximum possible score was 14 points. All applications received scores around 4 points (VR1: $M = 3.96$, $SD = 2.01$; VR2: $M = 4.00$, $SD = 1.96$; tutorial: $M = 4.16$, $SD = 1.70$). 

The \textit{Haptic} impression reflects the tactile perception of the VR application. This category has a maximum score of seven points. These normally distributed values were analyzed using a paired \textit{t}-test, which revealed no significant differences between conditions.

The \textit{Total Presence} score excludes sound and haptic according to the normative framework. It is based on the remaining five categories. No significant difference was observed between VR1 and VR2, $t(24) = 1.12$, $p = .272$, $95\% CI [-1.67, 5.67]$, $d = 0.22$, $95\% CI [-0.17, 0.62]$, with a mean difference of 2.00 points. However, both main applications differed significantly from the tutorial. VR1 was rated significantly higher than the tutorial, $t(24) = 3.12$, $p = .005$, $95\% CI [1.93, 9.51]$, $d = 0.62$, $95\% CI [0.19, 1.05]$, with an average difference of 5.72 points. Similarly VR2 scored significantly higher than the tutorial, $t(24) = 2.91$, $p = .008$, $95\% CI [2.24, 13.20]$, $d = 0.58$, $95\% CI [0.15, 1.00]$, with an average difference of 7.72 points.

%The total presence score excludes sound and haptics according to the normative framework. It is based on the remaining five categories and thus consists of 19 items in total. The tutorial received the lowest mean rating with 92.60 points (SD = 13.6), while VR1 achieved a higher mean rating of 98.32 (SD = 17.7), and VR2 obtained the highest score of 100.32  points (SD = 17.8). As with the previous measures, all total presence scores were within the normative range. No significant difference was observed between VR1 and VR2, $t(24) = 1.12, p = .272, 95\% CI [-1.67, 5.67], d = 0.22, 95\% CI [-0.17, 0.62]$, with a mean difference of -2 points. However, both main applications differed significantly from the tutorial. VR1 was rated significantly higher than the tutorial, $t(24) = 3.12, p = .005, 95\% CI [1.93, 9.51], d = 0.62, 95\% CI [0.19, 1.05]$, with an average difference of 5.72 points. Similarly and VR2 scored significantly higher than the tutorial, $t(24) = 2.91, p = .008, 95\% CI [2.24, 13.20], d = 0.58, 95\% CI [0.15, 1.00]$, with an average difference of 7.72 points.

%Summary
In summary, all three variants performed within the norm, with VR1 performing better in the \textit{Self-Evaluation of Performance} and VR2 performing better in terms of \textit{Realism}. Overall, the tutorial performed slightly worse than the other two variants, both in the \textit{Possibility to Examine}, as well as in the \textit{Total Presence}. In all other categories, the three variants performed similarly well.

\textit{Order Effects on Presence (VR1 and VR2):}
Since the study design required each participant to test all applications, scenarios arose in which a given VR application (VR1 or VR2) was used either before or after the other. To account for potential order effects, the data were further analyzed by splitting participants into two groups: \textit{VR1First} (VR1 tested before VR2) and \textit{VR1Second} (VR1 tested after VR2).

All seven presence-related categories were compared between these two groups. Although VR1 tended to receive slightly higher ratings when tested first, none of the differences reached statistical significance.

Two categories showed significant order effects for VR2. First, the realism rating was significantly higher when VR2 was tested first ($M = 38.7$, $SD = 8.96$) compared to when it was tested second ($M = 33.0$, $SD = 7.12$), as indicated by a Wilcoxon rank-sum test, $W = 114.5$, $p = .050$, with a moderate effect size, $r = 0.40$. Second, sound perception was significantly higher when VR2 was tested first ($M = 16.1$, $SD = 3.68$) compared to when it was tested second $11.2$ ($SD = 5.60$), as indicated by a Wilcoxon rank-sum test, $W = 117.5$, $p = .032$, with a moderate effect size, $r = 0.43$.

\subsection{Additional Application-Specific Questions}
%In addition to presence measures, further application-specific questions were posed regarding the impression of the spider, the design of the application, and its usability. Responses were collected using a Likert scale from 0 to 6, where 0 indicated no agreement and 6 indicated full agreement.

In addition to the presence questionnaire, separate application-specific questions were also compiled and collected. The visual impression of the spider held particular importance within the application, as it simulates the anxiety-inducing stimulus and serves as the central element of the behavioral approach test. The goal was to depict a realistic spider, and participants rated the realism on a scale from 0 = unrealistic to 6 = realistic. Although the spider model and animations were identical in VR1 and VR2, impressions were collected for both conditions to identify any potential differences between the scenarios. The results can be seen in ``Tab.~\ref{CostumResults}'', and show that the impressions were rated very similarly in both modalities.

%In VR1, ratings covered the entire scale range. On average, the spider's appearance was rated 3.64 (SD = 1.91). Animation quality of the spider received a similar average rating of 3.68 (SD = 2.14). In VR2, the full rating range was also covered. The visual quality of the virtual spider model received a mean rating of 3.36 (SD = 1.93), while the animations were rated with an average of 3.36 (SD = 1.98).

To gain deeper insight into perceptions of the spider’s behavior, an open-ended question was posed. Responses were qualitatively categorized into three groups: No changes, faster movement and slower movement.
%those expressing no desire for changes, those requesting more unpredictable and faster movements, and those preferring less movement.

In VR1, 15 out of 25 participants (60\%) reported that the spider's behavior matched their expectations. Four participants (16\%) expected a slower movement or even complete stillness, while the remaining six participants (24\%) expected faster or more unpredictable behavior. However, this expectation did not align for all participants with the preferred behavior of the spider. In fact, 16 participants (64\%) stated that the spider's behavior met their preferences. The predictability and consistent animation were particularly appreciated by most participants. Eight individuals (32\%) would have preferred no movement at all, while only one participant (4\%) desired faster movement, citing enhanced exposure effects as the reason. This pattern was mirrored in VR2. Although the responses of individual participants varied, the overall proportions remained identical. These variations, although based on different individuals, were not statistically significant ($W = 77$, $p = .3357$).

\begin{comment}

\begin{figure}[t]
            \centerline{\includegraphics[width=0.5\textwidth]{Figures/BATSB.png}}
        \caption{}
        \label{fig:SpiderBehaviorVR1}
    \end{figure}

\begin{figure}[t]
            \centerline{\includegraphics[width=0.5\textwidth]{Figures/Spider.png}}
        \caption{Expected and preferred behavior of the spider}
        \label{fig:SpiderBehaviorVR1}
    \end{figure}

\end{comment}

\begin{comment}
    
\begin{figure}[t]
            \centerline{\includegraphics[width=0.5\textwidth]{Figures/VRKURBELSPIDER.png}}
        \caption{}
        \label{fig:SpiderBehaviorVR1}
    \end{figure}

\begin{figure}[t]
            \centerline{\includegraphics[width=0.5\textwidth]{Figures/SpiderLaufVR.png}}
        \caption{}
        \label{fig:SpiderBehaviorVR2}
    \end{figure}

\end{comment}

The same questions regarding the spider's behavior were also posed during the real-life BATs to gain deeper insights into participants' expectations and anxiety. Due to missing data from one participant, the sample size for this part was $n = 24$.

In \textit{VIVO1}, only 25\% of participants reported that the spider’s behavior met their expectations. In contrast, 62.5\% stated they had expected the real spider to move more quickly or to move at all. A smaller proportion (12.5\%) expected less movement than was observed. 79.2\% stated that the spider's behavior matched their preferences. In contrast, 12.5\% desired even less movement and 8.3\% desired faster movement of the spider.
In \textit{VIVO2}, 33.3\% of participants expected the calm behavior of the real spider. However 66.7\% predicted a faster movement. Overall 87.5\% were satisfied with the calm behavior of the real spider. Only 8.3\% said they would have preferred a faster-moving spider, while 4.2\% preferred a slower-moving spider.

When comparing the categories between VR and vivo conditions, no significant differences were found regarding participants’ preferences ($W = 131$, $p = .1288$, $r = .281$). However, a significant difference with a high effect size emerged in participants’ expectations concerning the perceived spider behavior ($W = 33$, $p < .001$, $r = .535$).%, which is also reflected in the descriptive ``Fig. ~\ref{fig:SpiderBehaviorVR1}''.

\begin{comment}

\begin{figure}[t]
            \centerline{\includegraphics[width=0.5\textwidth]{Figures/REALKURBELBAT.png}}
        \caption{}
        \label{}
    \end{figure}

\begin{figure}[t]
            \centerline{\includegraphics[width=0.5\textwidth]{Figures/REALLAUFBAT.png}}
        \caption{}
        \label{}
    \end{figure}
  
\end{comment}

%\paragraph{Handling and Design Aspects}
Additional questions focused on the overall handling and audio of the application. The audio signal indicating the prompt for anxiety ratings was rated higher in VR1 ($M = 5.00$, $SD = 1.41$) than in VR2 ($M = 4.00$, $SD = 2.20$), based on a 6-point Likert scale. A statistical comparison revealed a significant difference between the two conditions ($W = 5$, $p = .00806$, $r = .502$). Also, the spoken explanation of the application was rated significantly different between these two conditions ($W = 0$, $p = .01298$,$r = .563$). Participants rated the audio explanation in VR1 with a mean of $5.52$ ($SD = 1.16$), compared to a mean of $4.80$ ($SD = 1.78$) in VR2, as seen in ``Tab.~\ref{CostumResults}''. 
This aspect was also compared with the real-world BAT. Although the instructions in the real conditions were delivered by the study coordinator rather than through the audio narration used in the VR applications, participants perceived the explanations in VR1 and VIVO1 similarly. There was no significant difference in rating between the two conditions ($W = 24$, $p = .904$, $r = .0608$). VIVO1 received a mean rating of $5.46$ ($SD = 0.69$).
In contrast, a significant difference was found between VR2 and VIVO2 ($W = 11.5$, $p = .03151$, $r = .474$). VIVO2 received a mean rating of $5.71$ ($SD = 0.69$), which is significantly higher than the rating given for VR2.
Overall, when comparing the aggregated BATs in virtuo to the aggregated BATs in vivo, no significant differences were observed ($W = 27.5$, $p = .0599$, $r = 0.208$).

The usability of the anxiety input via the controller slider received similar ratings, with no significant difference ($W = 63.5$, $p = .49$, $r = 0.188$). VR1 received a mean rating of $4.92$ ($SD = 1.45$), while VR2 was rated at $4.76$ ($SD = 1.19$). VIVO1 and VIVO2 were also rated very similarly, with an average rating of verbal anxiety ratings of $4.88$ ($SD = 1.12$) for VIVO1 and $4.92$ ($SD = 1.10$) for VIVO2.
Comparisons between VIVO1 and VR1 ($W=95$, $p = 1$, $r = .0961$), as well as VIVO2 and VR2 ($W = 47$, $p = .94$, $r = .0434$), indicate that the possibility to report anxiety was rated very similarly. Across both categories, the ratings were almost identical ($W = 198$, $p = .83$, $r = 0.096$).

% \usepackage{booktabs}
\begin{table}[]
\begin{tabular}{lcccc}
\toprule
\textbf{Category} & \textbf{VR1 Mean} & \textbf{VR1 SD} & \textbf{VR2 Mean} & \textbf{VR2 SD} \\ \midrule
Spider Optics     & 3.64           & 1.91            & 3.36           & 1.93            \\
Spider Animation  & 3.68           & 2.14            & 3.36           & 1.98            \\
Audio Feedback    & 5.00           & 1.41            & 4.00           & 2.20            \\
Explanation       & 5.52           & 1.16            & 4.80           & 1.78            \\
Anxiety Input     & 4.92           & 1.45            & 4.76           & 1.19            \\ \bottomrule \\
\end{tabular}
\caption{Custom Questionnaire Results}
\label{CostumResults}
\end{table}

\subsection{Physiological data}

%($p = .8311$, $r = 0.096$)
%TODO
%Nur Mean und STD
%P und R
%Condition 1 entfernen

%Vergleich mit fragebögen präsenz
%Correlation zwischen physiologie und Fragebögen
    %Einmal für die 3 BATs einzeln, einmal kombiniert
    %Vom custom Fragebogen nur Spinnenverhalten (wished und expected)

%Signal description
As mentioned earlier, heart rate, electrodermal activity (EDA), inter-beat interval and skin temperature were recorded using an Empatica E4. During preprocessing, EDA is split into the slower-changing tonic EDA, which reflects long-term factors such as stress, and the fast-changing phasic EDA, used to detect immediate reactions to specific stimuli. In a similar way the inter-beat interval is used to calculate two types of heart rate variability (HRV). Over short time windows, both the standard deviation of normal-to-normal heartbeat-intervals (HR\_SDNN), reflecting overall heart rate variability, and the root mean square of successive differences (HR\_RMSSD), indicating short-term reactions \cite{ReBAT}.  
After preprocessing, statistical measures, such as mean and standard deviation, were calculated. This was done separately for each BAT of every participant.

\begin{comment}
%Next, a Shapiro test was used to determine that the data follows a normal distribution. 
%Improve T-Test, include all values of the T-Test
Next, paired t-tests were used to determine if significant differences in the distributions were present between the different BATs, as well as the tutorial.
A heatmap of the significance values with FDR corrected p-scores is shown in figure \ref{fig:PhysHeatmap}. Significant differences were observed between condition 1 (questionnaires) and the other conditions, suggesting that condition 1 elicited distinct physiological responses compared to the rest. These differences were especially prominent in temperature-related features and heart rate/heart rate variability (HR/HRV) metrics.

Similarly, the tutorial, differed significantly from the four BATs, particularly in TEMP\_std, TEMP\_min, and HRV\_SDNN\_std, indicating modality-specific effects.

In contrast, comparisons within the cluster of the four BATs revealed fewer significant differences, suggesting more similar physiological profiles across these conditions. However, VR2 still showed consistent divergence from others, again particularly in TEMP and HRV measures, indicating a subtle but measurable distinction even within this group.

\begin{figure}[t]
    \centerline{\includegraphics[width=0.5\textwidth]{Figures/BAT Compare Heatmap.png}}
    \caption{T-Test between the different experimental conditions}
    \label{fig:PhysHeatmap}
\end{figure}
\end{comment}

\begin{figure}[t]
    \centerline{\includegraphics[width=0.5\textwidth]{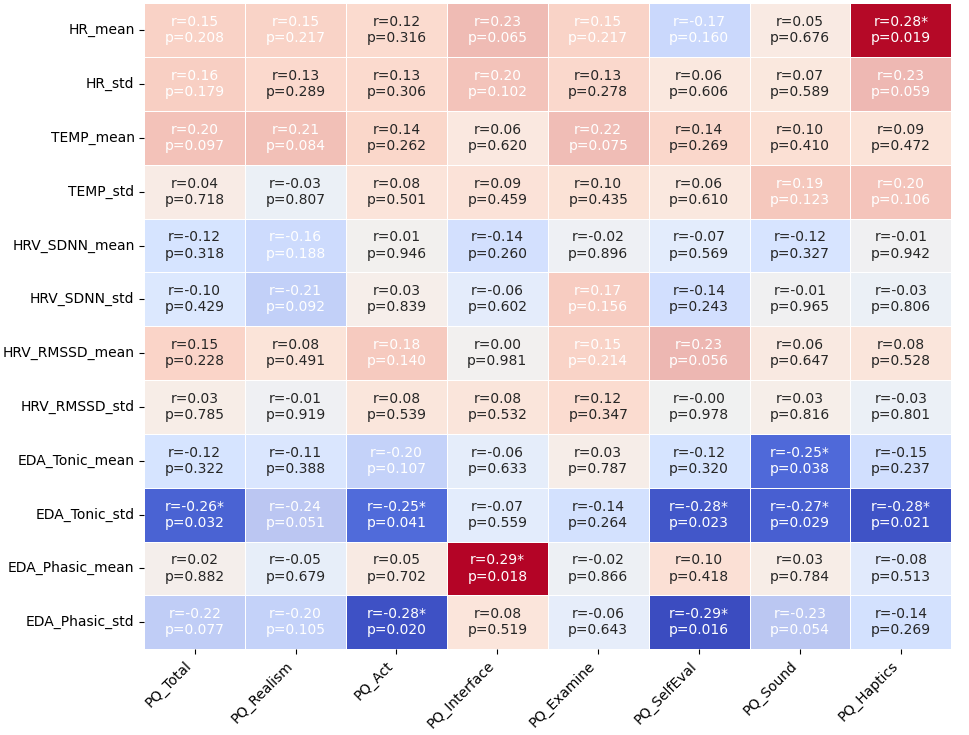}}
    \caption{Correlation between physiological signals and presence questionnaire}
    \label{fig:presence_combined}
\end{figure}

\begin{figure}[t]
    \centerline{\includegraphics[width=0.5\textwidth]{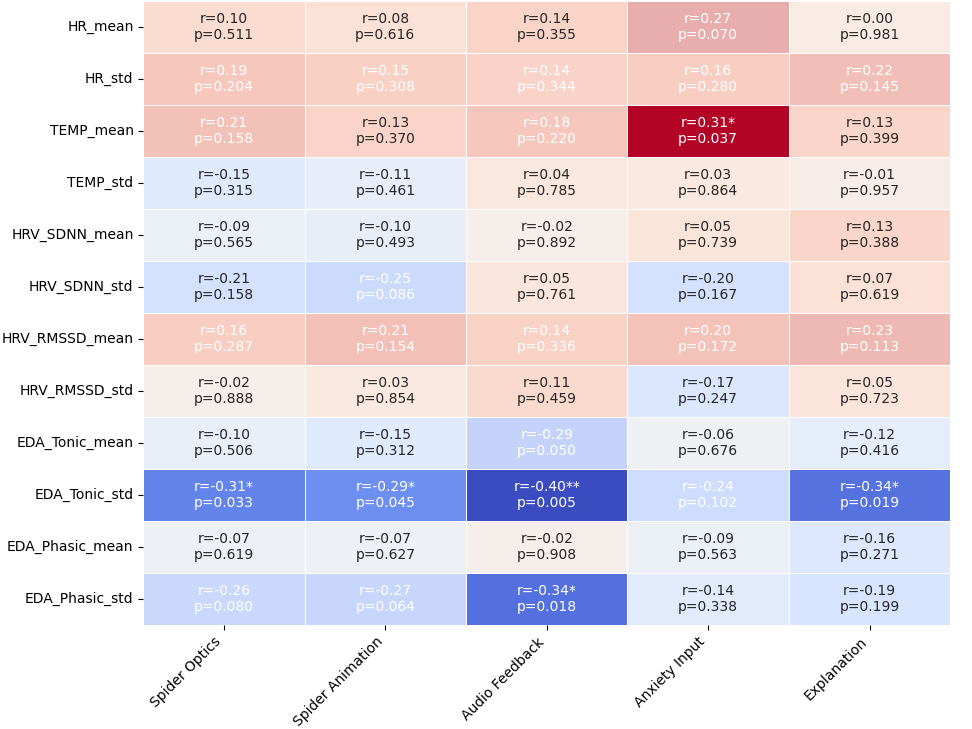}}
    \caption{Correlation between physiological signals and custom questionnaire}
    \label{fig:custom_combined}
\end{figure}

Next, a spearman correlation between the previously mentioned physiological signals and the individual questionnaire outcomes was calculated. The outcomes across all conditions are shown in ``Fig.~\ref{fig:presence_combined}'' and ``Fig.~\ref{fig:custom_combined}''. 

For the presence questionnaire, significant correlations were detected, particularly in the standard deviations of the EDA subtypes. While no significant correlations were detected for \textit{Realism} and \textit{Possibility to Examine}, for all other categories, at least one correlation reached statistical significance. For \textit{Possibility to Act}, a weak negative statistical significance was reached for both EDA\_Tonic\_std ($p = .041$, $r = -0.25$) and EDA\_Phasic\_std ($p = .020$, $r = -0.28$). For \textit{Quality of Interface}, a weak significance was reached in the correlation with EDA\_Phasic\_mean ($p = .018$, $r = 0.29$). \textit{Self-Evaluation of Performance} achieved significance with EDA\_Phasic\_std ($p = .016$, $r = -0.29$) as well as EDA\_Tonic\_std ($p = .023$, $r = -0.28$). Finally, \textit{Sound} has a correlation with with EDA\_Tonic\_std ($p = .029$, $r = -0.27$) and EDA\_Tonic\_mean ($p = .038$, $r = -0.25$), whereas \textit{Haptic} correlated with EDA\_Tonic\_std ($p = .021$, $r = -0.28$) and HR\_mean ($p = .019$, $r = 0.28$) . The \textit{Total Realism} correlated negatively with EDA\_Tonic\_std ($p = .032$, $r = -0.26$). 

For the custom questionnaire, data is only available for VR1 and VR2, as many questions focus on the spider, which is not present in the tutorial. Additionally, the questions about the expected and preferred behavior for the spider have been omitted, as the responses were too homogeneous, reducing the interpretability for correlations. Of the remaining categories, weak to moderate negative statistically significant correlations were reached between EDA\_Tonic\_std and \textit{Spider Optics} ($p = .033$, $r = -0.31$), \textit{Spider Animation} ($p = .045$, $r = -0.29$), \textit{Audio Feedback} ($p = .005$, $r = -0.40$) and the \textit{Explanation} of the task ($p = .019$, $r = -0.34$). In addition, a moderate correlation between EDA\_Phasic\_std and \textit{Audio Feedback} ($p = .018$, $r = -0.034$) was found. Lastly, a moderate positive correlation was reached between TEMP\_mean and the \textit{Anxiety Input} ($p = .037$, $r = -0.31$).

%Both the \textit{Spider Optics} and \textit{Spider Animation} correlated negatively with EDA\_Tonic\_std. The \textit{Expected Behavior} of the spider correlated highly significantly the mean of both HRV means, as well as the standard deviation of the heart rate. The \textit{Wished Behavior} has reached significance with RMSSD\_std and TEMP\_std. \textit{Audio Feedback} correlated with Phasic\_std and significantly correlated with Tonic\_std. The only positive correlation is between the rating for the \textit{Anxiety Input} and TEMP\_mean. Finally, the \textit{Explanation} in the application negatively correlated with the EDA\_Tonic\_std. 

\section{Discussion}

The evaluated data show notable differences between the individual BAT modalities and the subjective impressions of users. The analysis of presence provides important insights into the functionality and differences between the various BAT versions. All three scenarios fell within the normative range defined by the UQO Cyberpsychology Lab, indicating a sufficient level of presence and confirming the overall quality of the implementations. This can be seen as evidence that the variations, in their current form, are appropriate for use in anxiety assessment.

%Total Presence
However, the \textit{Total Presence} ratings reveal that environmental design plays a critical role in shaping the user experience. Both VR1 and VR2, which were developed for anxiety induction, received significantly higher total presence ratings compared to the tutorial, which focused solely on navigation training. This result is desirable, as the tutorial was intentionally designed as a low-pressure introduction, and participants could directly ask the study conductor questions. This instructional setting may have contributed to its lower presence rating.

%Realism and possibility to Examine
Environmental design may also have influenced the \textit{Realism} ratings. VR2, which replicated the layout of a real room, was rated as significantly more realistic than the other conditions. This suggests that both the realistic setting and the ability to physically walk through the space contributed to the immersive experience. 
The category \textit{Possibility to Examine} was rated significantly lower in the tutorial than in VR1 and VR2, even though all three applications technically offered the same interaction capabilities. The reduced number of elements in the tutorial room may have led to the lower perception in this category. Differences in the perceived \textit{Possibility to Act} were also noted. Although the same core interactions (e.g., anxiety input via controller) were implemented across conditions, participants felt more capable of acting in VR1 compared to the tutorial. This may be attributed to the room’s design and the presence of reactive elements such as an interactive avatar and a moving spider. There were no significant differences between VR1 and VR2, or between VR2 and the tutorial, even though VR2 provided fewer interaction elements.

%Self-Evaluation of Performance
The tutorial also received significantly lower ratings in the \textit{Self-Evaluation of Performance} category. One might have expected participants to feel most confident during the tutorial, as it was designed for skill acquisition and training. However, participants reported gaining more confidence in VR1 and VR2, possibly because those experiences involved meaningful interaction with the spider stimulus rather than general VR navigation.

%Sound, Haptics and Quality of Interface
Other aspects, such as sound, haptic, and interface quality, were rated similarly across all scenes. This was expected, as all three conditions used the same headset, haptic feedback, and audio elements. However, the audio of VR2 was rated slightly worse than VR1, with some participants noting difficulty localizing the audio in VR2. This is potentially due to the fact that, unlike in VR1, sound did not originate from a visible avatar. Another factor may be the increased physical demand in VR2, where participants navigated a real room while wearing the VR headset. This could have led to a reduction in auditory focus due to movement noise or cognitive load. Furthermore, the auditory issues observed in VR2 appear to have affected not only the perception of cues, but also the explanation of the application itself. In VR2, participants rated the instruction significantly lower than in the other BATs, suggesting that audio delivery in this condition may have been problematic. This points to specific auditory shortcomings in VR2 that may have hindered the user experience.

%Anxiety ratings
In contrast, the input of anxiety levels was perceived equally well across all conditions. No significant differences were found between verbal input in the real BATs and controller-based input in the virtual applications. This is a promising result, indicating that the digital anxiety input method can be reliably used in future studies without compromising comparability to real-world setups.

%Spider Visual
The perception of the virtual spider played a crucial role in the user experience across both VR variations. There were no significant differences between VR1 and VR2. This is an expected outcome, as the virtual spider’s appearance and animations were identical in both versions. However, the appearance only achieved a value just above the average, leaving room for improvement. Approximately 60\% of participants reported that the spider's appearance and behavior met their expectations and preferences. The slow and predictable movements contributed to a sense of realism while minimizing feelings of threat, yet the spider was still perceived as lifelike. This consistency was not mirrored in the real BAT conditions. Here, participants’ expectations did not align with the spider’s actual behavior. In most cases, the real spider remained still and passive, while many participants anticipated more movement or reactive behavior. In some trials, the spider did move more, but such behavior was unpredictable and uncontrollable, which complicates standardization. While the calm behavior of the real spider was generally preferred, it introduces variability that can affect anxiety measurement. Factors, such as the spider’s position within the box, sudden or absent movement, and even its size may influence both subjective perception and exposure outcomes. 

%Physiological data
The physiological data reveals a significant correlation between physiological data and presence values, as well as the appearance and animations of the spider. Consistent with previous research, like that by Frieden et al., EDA was found to be a particularly valuable signal when assessing presence, with most components of presence correlating with a more stable tonic EDA. However, no correlation between increased presence and increased EDA was detected in our work. Nevertheless, our findings show that using physiological signals can still be useful when evaluating the immersive experience. 
While these correlations are statistically significant, it is important to note that physiological signals are shaped not only by the presence itself, but also by emotional responses, such as anxiety. In a previous part of this project, we have successfully estimated the subjective anxiety ratings using the participants' physiological data \cite{ReBAT}. As such it is possible that the physiological response to anxiety is stronger than the response to presence, which might explain why no clear correlation between mean EDA and presence was determined in this study. For this reason, future studies into the physiological response to presence should focus on applications that do not elicit strong emotions.

%In fact, increased presence can amplify emotional reactions, meaning that stronger feelings of anxiety may arise as a direct consequence of heightened presence. This interplay suggests that emotional responses should be considered when interpreting physiological data, as they may act as mediating factors in how presence is experienced and measured.

%LIMITATIONS!
In addition to the aforementioned limitations, the study was subject to several general constraints. First, the sample primarily included women, as they are more likely to report fear of spiders compared to men. Second, some participants had previously taken part in a related study, which meant that they were already familiar with real spiders and may have responded differently as a result. Third, participants varied in their prior experience with virtual reality, and because each application could only be tested once, their impressions differed considerably. Finally, the relatively small sample size limits the extent to which the findings can be generalized.

\section{Conclusion}

%What we did:
This study examined different modalities of VR BAT in terms of presence and usability. It showed that although the different modalities were all within the norm in terms of presence, different aspects contribute to a good immersion. Additionally, the application received good feedback from the users, with the appearance and behavior of the spider, as well as the anxiety input method receiving good ratings.
%Comparison to other works
This work confirms existing knowledge about the general effectiveness of BATs in VR, such as by Mühlberger et al. \cite{Mühlberger} and our evaluation of psychometric measures \cite{}, and extends it by comparing different approach modalities and implementation specifications. 

%Lessens learned
Based on our evaluation, we propose the following practices: 
Firstly, in experiments containing both real and virtual environments, replicating the real environments appears beneficial. Additionally, introducing interactive elements, such as a virtual supervisor have a positive impact on presence, as well as ensuring that the localization of sound, such as instructions, is more intuitive. Furthermore, a clear goal, such as a confrontation with an anxiety-inducing stimulus also increase realism more, than simply learning the navigation or exploring. Lastly, increased freedom of movement appears to increase presence more than forcing the participant to remain in place. 

%Other advantages of VR
Other than these general lessons for increasing presence in VR applications, this work highlights some of the advantages of VR over classical approaches. The main advantage lies in the standardization, allowing for a fully controlled environment, that always produces the exact same conditions. This is especially important when it comes to the behavior of the spider, which can differ greatly between sessions in vivo. Other than standardization, VR also offers the opportunity for a fully controlled personalization. This could mean that the spider's behavior can be adjusted for each participant, while still remaining identical between sessions. 

In the future, the behavior of the virtual spider can be further improved, by adapting it to be more similar to the real spider. Rather than having the spider remain either fully unmoving, or in a constant, albeit slow, motion, it is beneficial to alternate between short active and passive phases of the spider. This increases the realism, while remaining in full control of the spider behavior, improving standardization. Additionally, future works can directly incorporate physiological signals, for example, to estimate anxiety ratings, and modify the behavior of the virtual spider according to the participant's response. 

Furthermore, VR glasses on the market are becoming increasingly affordable. An application like this could be transferred to commercially available portable glasses, which could then be used in psychotherapeutic practices. In this way, avoidance behavior could also be measured without using complicated setups, such as the crank construction, or a living spider.
Additionally, the application also offers the possibility to be modified for the fear of other specific small animals by exchanging the models and explanation audios, as well as the animations. This further opens up the field of research and provides new options for studying the individual perception of anxiety patients of specific phobias in a VR application.

\bibliographystyle{plain}
\bibliography{references}

\end{document}